\documentclass[epj]{svjour}

\input{epsf}
\input{amssymb.sty}

\begin{document}

\title{Interaction induced delocalization of two particles: 
large system size calculations and dependence on interaction 
strength}

\titlerunning{Interaction induced delocalization of two particles}

\author{Klaus M. Frahm}

\authorrunning{Klaus M. Frahm}

\institute{Laboratoire de Physique Quantique, UMR C5626 du CNRS, 
Universit\'e Paul Sabatier, F-31062 Toulouse Cedex 4, France, \\ 
\email{frahm@irsamc2.ups-tlse.fr}}

\date{September 14, 1998}

\abstract{
The localization length $L_2$ of two interacting particles 
in a one-dimensional disordered system is studied for very large 
system sizes by two efficient and accurate variants of the Green function 
method. The numerical results (at the band center) can be well described 
by the functional 
form $L_2=L_1[0.5+c(U)\,L_1]$ where $L_1$ is the one-particle localization 
length and the coefficient $c(U)\approx 0.074\,|U|/(1+|U|)$ 
depends on the strength $U$ of the on-site Hubbard interaction. 
The Breit-Wigner width or equivalently the (inverse) life time 
of non-interacting pair states is analytically 
calculated for small disorder and taking into account the energy 
dependence of the one-particle localization length. 
This provides a consistent theoretical explanation of the numerically 
found $U$-dependence of $c(U)$. 
} 

\maketitle

\section{Introduction}

\label{sec1}

The quantum eigenstates of non-interacting particles in a random potential 
are localized if the fluctuations of the potential (the disorder strength) 
are sufficiently strong (for a review see \cite{kramer1}). 
This phenomenon of Anderson 
localization is particularly well understood for one-dimensional or 
quasi one-dimen\-sional geometries where localization even persists for 
arbitrarily small disorder strength. For this case efficient numerical 
methods \cite{kramer1} and also powerful 
analytical theories in terms of the supersymmetric non-linear 
$\sigma$-model \cite{efetov1,guhr1} and the Fokker-Planck approach 
for the transfer matrix \cite{guhr1,been_rev1} are available. 

Dorokhov \cite{dorokhov1} and, recently, Shepelyansky \cite{shep1} 
considered the case of two interacting particles (TIP) moving in a 
one-dimensional 
random potential for which they predicted a strong enhancement of 
the localization length for the pair-states due to the interaction. 
While the results of Dorokhov are only valid for the case of a strongly 
attractive interaction confining both particles together, Shepelyansky 
also considered a local, attractive or repulsive Hubbard interaction. He 
claimed that among many states with both particles being localized far away, 
there are a few pairs-states where the typical distance between the 
particles is of the order of the one-electron localization length $L_1$ 
and the center of mass coordinate is characterized by a pair-localization 
length $L_2\gg L_1$. Mapping the original problem on a random band-matrix 
model superimposed with large diagonal elements, he found for the 
case $L_1\gg 1$ (length measured in units of the lattice spacing)
\begin{equation}
\label{eq1a}
L_2\sim \frac{U^2}{t^2}\,L_1^2
\end{equation}
where $U$ is the strength of the Hubbard interaction and $t$ is the value
of the hopping matrix element. The disorder strength enters through the 
value of $L_1$ (see below). This estimate has been confirmed 
by Imry \cite{imry1} using a different argument based on the Thouless 
scaling block picture \cite{thouless1}. A crucial role is played here 
by the spreading width $\Gamma$ (also called Breit-Wigner width) 
which is the energy scale over 
which unperturbed states are mixed due to the interaction. Imry 
identified this energy scale with a generalized Thouless energy defined 
as the sensitivity of the energy levels with respect to a change of the 
boundary conditions for a finite block of size $L_1$. The pair-localization 
length is then obtained by scaling theory as $L_2/L_1=\Gamma/\Delta_2$ 
where $\Delta_2\sim t/L_1^2$ is the two-particle level spacing in the 
finite block. Using an ergodic hypothesis for the one-particle eigenfunctions, 
one can estimate the spreading width by Fermi's golden rule 
$\Gamma\sim U^2/(t\,L_1)$ \cite{shep1,imry1,shep2} 
reproducing Eq. (\ref{eq1a}). 

First numerical studies in terms of finite size transfer matrix 
calculations \cite{frahm1} and exact diagonalization \cite{wein1} 
confirmed the strong enhancement due to the interaction. Borgonovi 
et al. \cite{borgonovi1} showed that the enhancement effect also 
appears in a related model of two interacting kicked rotors for which 
it is possible to determine directly the quantum time-evolution. 
Von Oppen et al. 
\cite{oppen1} introduced an efficient method to calculate the 
two-particle Green function and based on their numerical results they 
proposed the scaling relation $L_2/L_1\approx 0.5+0.054\,|U|\,L_1$ 
with a linear dependence on $U$ contradicting the estimate (\ref{eq1a}). 
This behavior was explained by Jacquod et al. \cite{jacquod2} who 
calculated analytically the spreading width for the limit of vanishing 
disorder. Resumming an infinite series of diagrams they obtained 
for energies close to the band center $\Gamma/\Delta_2\sim L_1\,
|U|/\sqrt{t^2+(U/4)^2}$. Therefore the physical arguments of Refs. 
\cite{shep1,imry1,shep2} did not contradict the results of 
\cite{oppen1} but the application of Fermi's golden 
rule corresponding to lowest order perturbation theory 
and the ergodic hypothesis 
appeared to be insufficient to determine $\Gamma$. 

It is worth mentioning that the topic also inspired considerable progress 
in the understanding \cite{jacquod1,fyodorov1,frahm2} of the random band 
matrix model with superimposed diagonal originally 
introduced and used by Shepelyansky \cite{shep1}. Furthermore, 
in Ref. \cite{frahm3}, a sophisticated random matrix 
model was proposed which works for arbitrary space dimension 
and takes properly both particle coordinates (relative and 
center of mass coordinate) into account. This model can be mapped onto 
an effective supermatrix non-linear $\sigma$-model and it is thus 
possible to explain features like a logarithmically suppressed diffusion 
or a logarithmically increasing pair size \cite{frahm3} found 
previously by Borgonovi et al. \cite{borgonovi1} 
for the pair-diffusive regime in $d>2$ dimensions where all states 
are delocalized. Subsequent work was concerned with 
the role of the level statistics \cite{wein2} and, very recently, 
with the fractal structure of the coupling matrix elements due to 
the interaction \cite{waintal1}. 

Despite the available evidence in favor for the enhancement effect the 
general situation is still not really clear, due to different proposals 
for the dependence of $L_2$ on $W$ and $U$ \cite{waintal1,pnomarev1,song1} 
and the claim of R\"omer et al. that the effect completely vanishes 
in the limit of infinite system size \cite{roemer1}. This claim, 
which was contested in \cite{frahm4}, is based on the finite size 
extrapolation of the localization length calculated by a transfer matrix 
method for finite square samples being put together to an infinite 
strip. 

In this work, we present numerical results (section \ref{sec2}) 
based on an exact and 
efficient variant of the Green function method introduced in 
Ref. \cite{oppen1} that allows to treat rather large system 
sizes, i.~e.  $N\ge 1000$. This is indeed important for small disorder 
values in order to perform an 
accurate finite size extrapolation. We furthermore present a second 
variant which consists of the recursive Green 
function technique applied to an effective band matrix Hamiltonian as 
considered in \cite{oppen1}. In this approach one can indeed take 
the limit $N\to\infty$ and the results we find are 
consistent with those of the finite size extrapolation of the first 
variant. The issue of an accurate variant of the Green function method 
is actually of considerable interest since R\"omer et al. \cite{roemer1} 
had questioned the original results of von Oppen et al. due to a certain 
approximation 
applied in the original approach of Ref. \cite{oppen1}. We find in our 
calculations qualitative agreement with those results concerning the 
strong enhancement of the localization length $L_2$ and the dependence 
on $L_1$. However, we find nevertheless a quantitative difference 
concerning the dependence on the interaction strength $U$ which 
is only linear for sufficiently small $U$. To understand this, 
we reconsider the issue of the determination of the Breit-Wigner width 
$\Gamma$ for small disorder (section \ref{sec3}). Improving the 
$\Gamma$ estimate of Ref. \cite{jacquod2}, we can indeed explain the 
modified $U$ dependence. 

Very recently, we learned of related relevant work 
\cite{leadbeater1,song2} in which the TIP Green function was evaluated 
by a decimation method for system sizes up to $N=251$ \cite{leadbeater1} 
or $N=300$ \cite{song2}.

\section{Numerical Green function approach}

\label{sec2}

We consider two particles in a disordered system interacting via 
a local Hubbard-interaction and characterized by the following 
tightbinding Hamiltonian,
\begin{eqnarray}
\nonumber
H&=&-t\sum_{x,y}\Bigl(|x+1,y><x,y|+|x,y+1><x,y|+\mbox{h.c.}\Bigr)\\
&&+\sum_{x,y}\Bigl(\varepsilon(x)+\varepsilon(y)+U\,\delta_{x,y}\Bigr)
|x,y><x,y|\ .
\label{eq1}
\end{eqnarray}
$x$ and $y$ denote the positions of the first and the second particle, 
respectively. $t$ is the strength of nearest neighbor coupling matrix element 
which we put to unity in the following and $U$ is the value of the on-site 
interaction. The disorder energies are random, i.~e. 
$\varepsilon(x)\in[-W/2,\,W/2]$ with $W$ being the disorder strength. 
At vanishing interaction $U=0$ the one-particle eigenstates (at a one-particle 
energy $E=0$) are localized with the localization length: 
$L_1\approx 105/W^2$ \cite{kramer1}. In this work we do not discuss the 
particular effects of symmetric or anti-symmetric two-particle states 
(Bosons or Fermions). The on-site Hubbard interaction only acts on the subspace 
of symmetric states and our results apply therefore to the case of Bosons. 
However, for the actual calculations and the representation, we find it more 
convenient to keep all states. 

To determine the two-particle 
localization length, we consider as in Ref. \cite{oppen1} the 
two-particle Green function. Since we are interested in the coherent 
propagation the particles being close, we determine only the 
Green function matrix elements of doubly occupied states $|xx>$, 
\begin{equation}
\label{eq2}
g_{xy}=<xx|\,(E-H)^{-1}\,|yy>\ .
\end{equation}
A priori, for a finite system of size $N$, the matrix inverse in (\ref{eq2}) 
has to be evaluated for a $N^2\times N^2$ matrix. Fortunately, von Oppen et 
al. \cite{oppen1} have shown that this problem can be reduced to an 
effective Green function on an $N$-dimensional space because the interaction 
operator is proportional to the projector on the space of doubly occupied 
states. The matrix $g$ in (\ref{eq2}) can be calculated 
\cite{oppen1} from an $N\times N$-matrix equation 
\begin{equation}
\label{eq3}
g=g_0\,\frac{1}{1-g_0\,U}\quad,\quad\mbox{where}\quad g_0=g\Big|_{U=0}\ . 
\end{equation}
The matrix $g_0$ is given in terms of 
the one-particle eigenstates $\varphi_\alpha(x)$ and the one-particle 
energies $E_\alpha$ via,
\begin{equation}
\label{eq4}
(g_0)_{xy}=\sum_{\alpha,\beta}\varphi_\alpha(x)\,\varphi_\beta(x)\,
\frac{1}{E-E_\alpha-E_\beta}\,\varphi_\beta(y)\,\varphi_\alpha(y)\ . 
\end{equation}
The two-particle localization length $L_2$ is determined by the exponential 
decay $g_{x_0,x}\sim\exp(-|x-x_0|/L_2)$ corresponding to 
\begin{equation}
\label{eq5}
\frac{1}{L_2}=-\lim_{N\to\infty}\frac{1}{N}\left\langle
\ln\left|\frac{g_{x,x+N}}{g_{x,x}}\right|\right\rangle\ .
\end{equation}
The ensemble average is performed over different disorder realizations 
and for practical purposes also over some initial sites $x$ 
close to one boundary. The extra denominator $g_{x,x}$ in (\ref{eq5}) is not 
relevant in the limit $N\to\infty$ but provides a considerable improvement 
if (\ref{eq5}) is evaluated for finite $N$. For vanishing interaction, 
we expect according to (\ref{eq4}) $L_2(U=0)\approx L_1/2$ \cite{oppen1}. 

In Ref. \cite{oppen1}, Eq. (\ref{eq3}) was evaluated for a finite 
system by employing two approximations. First, von Oppen et al. omitted 
the first factor $g_0$ and, second, they did not evaluate the full matrix 
$g_0$ but only a sufficiently large band on which they applied 
the recursive 
Green function technique \cite{mackinnon1,hucke1} for the matrix inverse in 
(\ref{eq3}). Since $g_0$ is indeed a band matrix of width $\sim L_1$
both approximations seem to be well justified provided $L_1<L_2/2$. 
However, the validity of the corresponding results was seriously questioned by 
R\"omer et al. \cite{roemer1} due to these approximations and, furthermore, 
the limit of very small (vanishing) interaction cannot accurately be 
studied within this approach. 

We have evaluated (\ref{eq3}) exactly without any approximations. 
For this we note two important points concerning the numerical precision 
and the efficiency. First, the multiplication of the band matrix 
$g_0$ with the matrix inverse in (\ref{eq3}) requires that the 
{\em relative} error of the exponentially small matrix elements of $g_0$ 
far away from the diagonal is small \cite{foot1}. 
Otherwise, Eq. (\ref{eq5}) provides 
incorrect results for $L_2$. This in turn requires that the exponential 
tails of the $\varphi_\alpha(x)$ are accurate over the whole length scale 
$x=1,\ldots,N$. According to this, we have determined the one-electron 
eigenstates by the method of inverse vector iteration \cite{schwartz1} 
which provides the required accuracy by sufficiently increasing the number 
of iterations. The second point concerns the efficiency. Here 
the matrix inverse in (\ref{eq3}), which costs of the order 
of $N^3$ operations, is actually not the limiting factor. This is due to 
the necessary evaluation of $g_0$. The naive application of 
(\ref{eq4}) already costs of the order of $N^4$ operations. 
Even though this number can be reduced by exploiting the exponential 
decay of the $\varphi_\alpha(x)$ this does not yield any significant 
improvement for small disorder values when $L_1\sim 50-100$. 
Fortunately, it is possible to determine $g_0$ exactly with $N^3$ operations. 
For this we rewrite (\ref{eq4}) in the form
\begin{equation}
\label{eq6}
(g_0)_{xy}=\sum_\alpha\varphi_\alpha(x)\,G_{xy}^{(1)}(E-E_\alpha)\,
\varphi_\beta(y)\ ,
\end{equation}
where $G_{xy}^{(1)}(E)$ is the one-particle Green function at energy $E$ 
that can efficiently be determined by $N^2$ operations due to 
the tridiagonal form of the one-electron Hamiltonian. Since this has to be 
done for $N$ different energies $E-E_\alpha$, Eq. (\ref{eq6}) provides 
an algorithm with only $N^3$ operations. 

We have used two variants of the Greens function method. The 
first is based on a finite size extrapolation (FSE) to determine 
the limit $L_2=\lim_{N\to\infty}L_2(N)$. For this, we have 
calculated the ensemble averaged inverse localization length $L_2^{-1}(N)$ 
for different system sizes using the exact projected 
Green function $g$ as given in (\ref{eq3}). The limit for $N\to\infty$ 
has been determined by the linear fit in $1/N$ of the {\em inverse} 
localization length:
\begin{equation}
\label{eq7}
\frac{1}{L_2(N)}\approx\frac{1}{L_2}+\frac{C}{N}\ .
\end{equation}
This ansatz for the finite size extrapolation is highly suggestive 
from Eq. (\ref{eq5}). Assuming, that the typical value of $g_{x,x+N}$ does not 
depend on $N$ if $N\ll L_2$, we see that (\ref{eq7}) reproduces 
both limits $N\ll L_2$ and $N\gg L_2$. 

\begin{figure}
\epsfxsize=3in
\epsffile{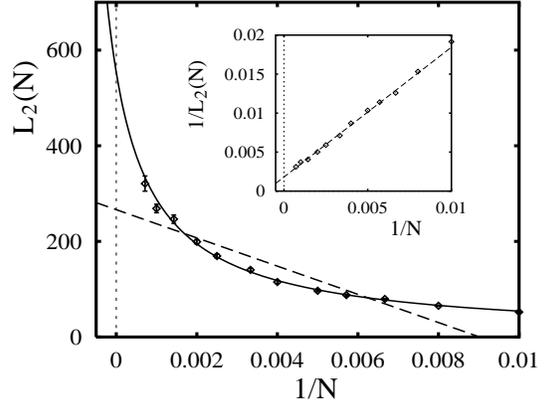}
\medskip
\caption{Finite size two-particle localization length $L_2(N)$ for 
$U=1.5$, $E=0$ and $W=1.0$ versus $1/N$ ($100\le N\le 1400$). The full curve 
corresponds to the fit $L_2^{-1}(N)=L_2^{-1}+C/N$ and the dash line 
corresponds to the fit $L_2(N)=L_2+\tilde C/N$. The insert shows $1/L_2(N)$ 
versus $1/N$ with the corresponding linear fit.}
\label{fig1}
\end{figure}

The quality of the fit is indeed confirmed by the explicit numerical 
values as can be seen in Fig. \ref{fig1}. For $W=1.0$ ($L_1=105$), $E=0$, 
$U=1.5$, and $100\le N\le 1400$, the fit (\ref{eq7}) works quite well 
while the direct fit $L_2(N)=L_2+\tilde C/N$ is very poor for the 
considered range of $N$ values. Of course, for $L_2\ll N$ 
both extrapolation schemes are equivalent. However, for the case of 
Fig.~\ref{fig1}, $L_2$ and $N$ are 
comparable and the choice of the correct method is crucial. 

\begin{figure}
\epsfxsize=3in
\epsffile{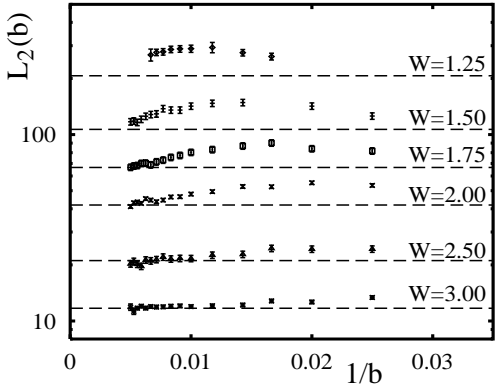}
\medskip
\caption{Localization length $L_2(b)$ (on a logarithmic scale) for $U=1$ and 
$E=0$ obtained by the second method 
[recursive Green function method applied on (\ref{eq3})] versus the inverse 
block size $1/b$ ($40\le b\le 200$). The dashed lines 
correspond to the values of $L_2$ obtained by the first method (finite 
size extrapolation).}
\label{fig2}
\end{figure}

To obtain an independent verification of the extrapolation scheme 
(\ref{eq7}), we have also used a second method 
which permits to determine $L_2$ directly for quasi 
infinite systems. For this, as in Ref. \cite{oppen1}, we have omitted 
the first factor in (\ref{eq3}) and replaced in the denominator of the 
second factor the matrix $g_0$ by a block-tridiagonal matrix of 
block size $b$. 
This matrix can be viewed as an effective Hamiltonian \cite{oppen1} and 
it is possible to perform the matrix inverse of (\ref{eq3}) by 
the recursive Green function (RGF) technique \cite{mackinnon1,hucke1}. 
In contrast to von Oppen et al., we have applied this method for quasi 
infinite system size as described in Ref. \cite{mackinnon1}. The limit 
(\ref{eq5}) can be directly evaluated due to the self averaging 
behavior of the inverse localization length. The non-vanishing blocks 
of $g_0$ are determined in a local approximation, i.e. to 
calculate $(g_0)_{xy}$ for $x,y\in\{x_0-b+1,\ldots,x_0+b\}$ we apply 
Eq. (\ref{eq6}) for a finite system of size $N_c>2b$ containing the 
sites $x\in\{x_0-N_c/2+1,\ldots,x_0+N_c/2\}$. This works if $N_c$ and $b$ 
are sufficiently large compared to $L_1$ because 
the matrix elements $(g_0)_{xy}$ do not depend on $N_c$ 
if the sites $x,y$ are far away from the boundaries. For each iteration 
step of the recursive Green function procedure the disorder realization 
on the $N_c$ sites is shifted by $b$ sites and $g_0$ has to be recalculated 
by (\ref{eq6}). We have chosen $N_c=3b$ and calculated $L_2(b)$ for 
different values of the block sizes in the interval $40\le b\le 200$. 
Here, we expect a much faster (exponential) convergence of $L_2$ as the 
ratio $b/L_1$ becomes larger than unity. In Fig.~\ref{fig2}, we compare 
for $U=1$ and different disorder values, $1.25\le W\le 3.00$, the 
localization length $L_2(b)$ with the values obtained 
from the FSE method. 

We find overall agreement between both methods and $L_2(b)$ indeed 
coincides with $L_2$ for $b\ge b_c\approx 5\,L_1$. For small disorder 
values it is quite difficult to arrive at this regime. For $b<b_c$ 
the values of $L_2(b)$ are typically larger than the values obtained by the 
first method. 
To our knowledge, the approach described above is indeed the first method 
to determine directly the TIP localization length 
for quasi infinite system size without the side effects of 
a bag-interaction \cite{shep1,frahm1}. This is possible, because the 
cutoff is applied on an effective Hamiltonian and the neglected 
matrix elements are indeed exponentially small if the 
block size $b$ is sufficiently large. 
The results shown in Fig. \ref{fig2} provide therefore an 
additional confirmation of the validity of the above discussed 
extrapolation scheme. 

\begin{figure}
\epsfxsize=3in
\epsffile{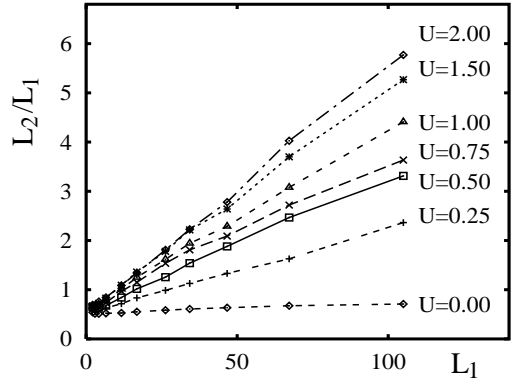}
\medskip
\caption{Enhancement factor $L_2/L_1$ as a function of $L_1=105/W^2$. 
$L_2$ is extrapolated as in (\ref{eq7}) using system sizes up to 
$N\le 1000$. 
The considered disorder values are 
$W=1,\,1.25,\,1.5,\,1.75,\,2,\,2.5,\,3,\,4,\,5,\,6,\,7$. 
}
\label{fig3}
\end{figure}

For a systematic study of the dependence on $W$ and $U$, we used the 
first variant based on the FSE scheme (\ref{eq7}) 
which appears to be more efficient, especially for 
large values of $L_1$. For the scope of this paper, we studied the 
band center $E=0$ where the localization properties are symmetric 
with respect to the sign of $U$. We considered for the 
disorder values $1.0\le W\le 7.0$ 
and interaction strengths $0.0\le U\le 2.0$ 
at least system sizes up to $N=500$ and for $W\le 1.75$ even sizes up to 
$N=1000$. (For $W=1.0$ and $U=1.5,\,2.0$, we have also calculated 
two data-points with $N=1400$.) Most of the data points (for the finite 
size values $L_2(N)$) were determined with a relative error 
smaller than $2\ \%$. For the largest system sizes and smallest disorder values 
the relative error is $3-3.5\ \%$. 
To verify the scaling relation $L_2/L_1\approx 0.5+0.054\,|U|\,L_1$ 
suggested by von Oppen et al. \cite{oppen1}, we show in Fig. \ref{fig3} 
the ratio $L_2/L_1$ as a function of $L_1$ where $L_2$ has been obtained 
by FSE from $L_2(N)$. 

The linear behavior in $L_1$ is qualitatively indeed confirmed but the errors 
for the smaller disorder values do not allow to exclude a behavior of the type 
$(L_2/L_1-0.5)\propto L_1^\alpha$ with $\alpha<1$. A corresponding fit 
indeed gives 
$\alpha\approx 0.9$ but this depends also on the chosen offset $0.5$ 
(the fit with the offset $0.55$ gives $\alpha\approx 1.0$). 
We mention that the slight deviations from the linear behavior can 
also be well described by an ansatz of the type 
$(L_2/L_1-0.5)\propto L_1/\ln(C\,L_1)$ suggested 
by Borgonovi et al. \cite{borgonovi1}. 
However, since the precision of the data does not permit to distinguish 
significantly between this and the linear behavior, we do not enter into 
more details here. 
For $U=0$, we confirm the previous observation \cite{song1,leadbeater1} 
of a slight enhancement $L_2(U=0)/L_1\approx 0.5-0.7$ which is presumably 
due to the energy average in Eq. (\ref{eq4}) \cite{song1,leadbeater1}.

\begin{figure}
\epsfxsize=3in
\epsffile{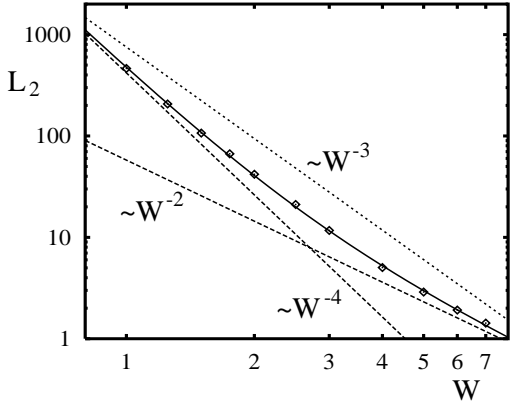}
\medskip
\caption{Dependence of $L_2$ on $W$ for $U=1$ (double log-scale). 
The data-points are the same as in Fig. \ref{fig3}. The full curve 
is the fit $L_2=L_1\,(0.55+0.038\,L_1)$ with $L_1=105/W^2$. The 
two dashed lines correspond to the limiting cases 
$L_2=0.038\,L_1^2\propto W^{-4}$ (or $L_2=0.55\,L_1\propto W^{-2}$) 
for $L_1\gg 10$ ($L_1\ll 10$). The dotted line corresponds the 
behavior $\propto W^{-3}$ for comparison. }
\label{fig4}
\end{figure}

In Fig. \ref{fig4}, we also show the dependence of $L_2$ on the disorder 
strength $W$ (for $U=1$ and $E=0$). Previously, Song et al. \cite{song1}
found a behavior $L_2\propto W^{-2.9}$ and as we can see the overall slope 
in Fig. \ref{fig4} is indeed comparable to this behavior. However, we find 
for small and large $W$ values significant deviations due to the curvature 
in the curve of $\ln(L_2)$ versus $\ln(W)$. This is due to the constant term 
in the above scaling relation. Actually, the data can be extremely 
well fitted by 
$L_2\approx L_1(0.55+0.038\,L_1)$ for the whole interval $1.0\le W\le 7.0$ 
and one indeed finds the asymptotic 
behavior $L_2\propto W^{-4}$ for small $W$ ($L_1\gg 10$) and 
$L_2\propto W^{-2}$ for larger values of $W$ ($1<L_1<10$). 

\begin{figure}
\epsfxsize=3in
\epsffile{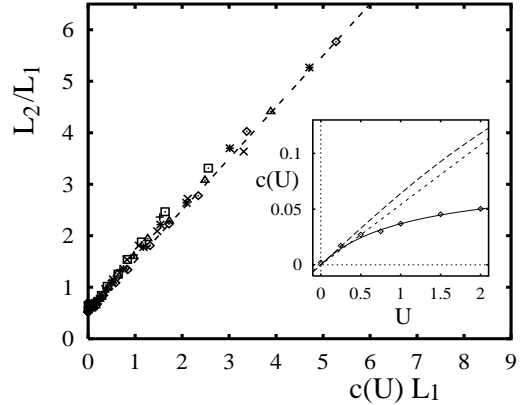}
\medskip
\caption{$L_2/L_1$ versus the scaling parameter $c(U)\,L_1$ for 
the data of Fig. \ref{fig3} (points) where the quantity $c(U)$ 
is given by Eq. (\ref{eq23}) with $A=0.028$ and $k_c=0.25$ 
(see text). The symbols for the different $U$ values are the 
same as in Fig. \ref{fig3}. 
The dashed line corresponds to $L_2/L_1=0.5+c(U)\,L_1$. 
The insert shows $c(U)$ as a function of $U$. The data points are 
the values obtained as the slope in the fit $L_2/L_1=a+c(U)\,L_1$. 
The full line is the analytical expression (\ref{eq23}) 
(same values for $A$ and $k_c$ as above), 
the dotted line is the linear dependence $0.054\,U$ found 
in Ref. \cite{oppen1} and the 
dashed line is the dependence suggested in Ref. \protect\cite{jacquod2}.} 
\label{fig5}
\end{figure}

To extract the dependence on $U$, 
we determined the slope $c(U)$ in the linear fit $L_2/L_1=a+c(U)\,L_1$ 
which is compared in Fig. \ref{fig5} with the numerical data. 
The slope $c(U)$ itself is shown in the insert of 
Fig. \ref{fig5} as a function of $U$. 
Apparently, the $U$ dependence is not linear for the whole interval 
$0.0\le U \le 2.0$. This linear behavior was observed by v. Oppen et al. 
\cite{oppen1} for $U \le 1.0$ where the 
discrepancy is still quite moderate. At $U=1.0$ their values are 
about $40 \%$ larger than ours. 
We believe that this is due to finite size effects and the 
applied approximations.

Also the estimate $c(U)\propto |U|/\sqrt{1+(U/4)^2}$ based on 
the analytical calculation of the Breit-Wigner width for $W=0$ 
\cite{jacquod2} only agrees with the numerical data for 
sufficiently small $U$. 
In the next section, we will try to explain this disagreement and 
reconsider the determination of the Breit-Wigner width. 

\section{Breit-Wigner Width and $U$ dependence}

\label{sec3}

The delocalization effect of two interacting particles is related to the 
finite life time $\tau=\Gamma^{-1}$ of the product states $|\alpha\beta>$ with 
$<x_1\,x_2|\alpha\beta>=\varphi_\alpha(x_1)\,\varphi_\beta(x_2)$ 
\cite{shep1,imry1,shep2}. The interaction gives rise to transitions 
$|\alpha\beta>\to|\gamma\delta>$ which can be viewed as random hops of 
typical size $L_1$. For short time scales when quantum interference 
effects can be neglected one therefore obtains a diffusive dynamics 
\cite{shep1,shep2} with the diffusion constant 
$D\sim L_1^2\tau^{-1}=L_1^2\Gamma$. Following a general argument 
developed in Ref. \cite{chirikov1} and applied to the TIP case 
in Refs. \cite{shep1,shep2}, one can estimate the localization length 
due to quantum interference effects. According to this the 
classical diffusive behavior is only valid for time scales smaller than 
the Heisenberg time corresponding to a wave packet of width $\sqrt{D t}$, 
i.~e.
\begin{equation}
\label{eq9}
t<t_H(t)=\nu_{\rm eff}\,\sqrt{Dt}\ .
\end{equation}
Here $\nu_{\rm eff}$ is the density of states per length such that 
$(L\nu_{\rm eff})^{-1}$ is the level spacing in a block of size $L$. 
For the TIP, we have $\nu_{\rm eff}\approx \nu_2\,L_1$ with 
$\nu_2(E)$ being the energy dependent two-particle density of states. 
$\nu_2(E)$ is proportional to the inverse bandwidth 
$1/(8+2W)\sim 1$ with a logarithmic singularity at the band center 
(for $W=0$). The density $\nu_{\rm eff}$ corresponds to the number of 
well coupled pair states with the same center of mass coordinate but 
with different relative coordinates (up to a maximal value $\sim L_1$). 

At $t\approx t_c=\nu_{\rm eff}^2 D$, when the condition (\ref{eq9}) 
ceases to be valid, the discrete energy spectrum 
can be resolved and localization with a localization length 
$L_2\sim\sqrt{D t_c}\sim \nu_{\rm eff}\,D$ sets in. This general relation 
for quasi 1d systems has also been obtained in a more rigorous way using the 
supersymmetric non-linear $\sigma$-model \cite{efetov1}. In view of this, the Breit-Wigner width $\Gamma$ and 
$L_2$ are related by
\begin{equation}
\label{eq10}
L_2\sim\nu_{\rm eff}\,D\sim L_1^3\,\Gamma\nu_2.
\end{equation}
Jacquod et al. \cite{jacquod2} determined $\Gamma$ using diagrammatic 
perturbation theory (in $U$) for the case of 
vanishing disorder, $W=0$, and finite system size $N$. They argued that 
one obtains a good estimate of $\Gamma$ for finite $W$ by replacing 
$N$ with $L_1$ due to the ballistic dynamics in $1d$ for length scales up 
to $L_1$. For energies close to the band center and moderate interaction 
strengths $U \lesssim 1$, their result reads
\begin{equation}
\label{eq11}
L_2\sim L_1^2\frac{|U|}{\sqrt{1+(U/4)^4}}\ .
\end{equation}
As we can see in the insert of Fig. 5 our numerical data agrees with 
this behavior only for very small values of $U$. 
We therefore feel that it is justified to reconsider 
the issue of the Breit-Wigner width which can be calculated from the 
energy dependent local density of states
\begin{equation}
\label{eq12}
\rho_{\alpha\beta}(E)=-\frac{1}{\pi}\mbox{Im}\,
<\alpha\beta|\,(E+i0-H)^{-1}\,|\alpha\beta>\ ,
\end{equation}
with $|\alpha\beta>$ as above. Using Schur's formula to perform the 
matrix inverse, we rewrite (\ref{eq12}) as 
\begin{eqnarray}
\label{eq13}
\rho_{\alpha\beta}(E)&=&-\frac{1}{\pi}\mbox{Im}\,
\Bigl[E+i0-E_\alpha-E_\beta-\\
\nonumber
&&-<\alpha\beta|\,\hat U\,|\alpha\beta>
+\Gamma_{\alpha\beta}/2\Bigr]^{-1}\quad,\\
\label{eq14}
\Gamma_{\alpha\beta}&=&\Gamma_{\alpha\beta}^{(0)}+
i\,\Gamma_{\alpha\beta}^{(1)}\\
\nonumber
&=&-2<\alpha\beta|
\,\hat U\,(E+i0-\tilde H)\,\hat U\,|\alpha\beta>\quad.
\end{eqnarray}
Here $\hat U=U\sum_x |xx><xx|$ is the interaction operator and 
$\tilde H=\tilde P\,H\,\tilde P$ with the projector 
$\tilde P=1-|\alpha\beta><\alpha\beta|$. Eq. (\ref{eq13}) corresponds 
to the Lorentzian or Breit-Wigner form of the local density of states 
provided that the energy dependence of $\Gamma_{\alpha\beta}$ is weak. 
The imaginary part $\Gamma_{\alpha\beta}^{(1)}$ is the width of 
the Lorentzian. In the following, we replace $\tilde H$ by $H$, and 
we first evaluate (\ref{eq14}) for the case of vanishing disorder. 
For this we only need the projected Green function (\ref{eq2}) due to 
the appearance of $\hat U$ in (\ref{eq14}). In view of 
Eq. (\ref{eq3}), we first determine
\begin{eqnarray}
\label{eq15}
(g_0)_{xy}&=&<xx|\,(E+i0-H_0)^{-1}\,|yy>\\
\nonumber
&\approx&
\frac{1}{2\pi}\int_{-\pi}^\pi dk\ e^{ik(x-y)}\tilde g_0(k)
\end{eqnarray}
with
\begin{eqnarray}
\nonumber
\tilde g_0(k)&=&\frac{1}{2\pi}\int_{-\pi}^\pi dq
\,\frac{1}{E+i0+2[\cos(q)+\cos(k-q)]}\\
\label{eq16}
&=&-i\frac{1}{\sqrt{4\cos^2(k/2)-E^2}}\ .
\end{eqnarray}
The Green function at $W=0$ and $U\neq 0$ is then given by
\begin{equation}
\label{eq17}
g_{xy}=\frac{1}{2\pi}\int_{-\pi}^\pi dk\ e^{ik(x-y)}
\ \frac{\tilde g_0(k)}{1-U\,\tilde g_0(k)}\ .
\end{equation}
 From this and Eq. (\ref{eq14}), we obtain
\begin{equation}
\label{eq18}
\Gamma_{\alpha\beta}=-2U^2\,\sum_{x,y}\varphi^*_\alpha(x)\,
\varphi^*_\beta(x)\,g_{xy}\,\varphi_\alpha(y)\,
\varphi_\beta(y)\ .
\end{equation}
Inserting the plane wave eigenstates for $\varphi_{\alpha}$ or 
$\varphi_\beta$, we exactly recover the result of Ref. \cite{jacquod2} 
for the Breit-Wigner width. This shows that the diagrammatic approach of 
\cite{jacquod2} is equivalent to our above approximations 
(replacing $\tilde H$ by $H$ and continuum limit for $k$). 

The generalization to the disordered case essentially gives rise to 
two modifications. Using diagrammatic perturbation theory in the disorder, 
one can first evaluate the average of the Green functions (\ref{eq15}) and 
(\ref{eq16}) which amounts to the replacement $E+i0\to E+i\gamma$ where 
$\gamma$ is determined by a Dyson equation. For weak disorder one finds 
$\gamma\sim W^2$ (with eventual logarithmic corrections at the band 
center). In the following discussion, we neglect the effect 
of this small $\gamma$ which essentially regularizes (\ref{eq16}) at the 
singularity. The second, more important, modification due to finite 
disorder concerns the eigenfunctions $\varphi_\alpha(x)$ in Eq. (\ref{eq18}). 
Here the energy dependence of the one particle localization length plays 
an important role. To see this, we use the toy ansatz
\begin{equation}
\label{eq19}
\varphi_\alpha(x)\approx\frac{1}{\sqrt{L_{1,\alpha}}}\,
e^{-|x-x_\alpha|/L_{1,\alpha}+i\,k_\alpha\,x}
\end{equation}
with $E_\alpha=-2\cos k_\alpha$ and an energy dependent localization length 
$L_{1,\alpha}\approx L_1\,\sin^2(k_\alpha)$ \cite{kramer1}. This ansatz 
essentially corresponds to a particle moving ballistically 
with a well defined momentum inside the localization domain around $x_\alpha$. 
This is indeed reasonable because in one dimension the mean free path is of 
the same order as the localization length. However, 
the momentum has to be larger than its typical uncertainty, i.~e. 
$|k_\alpha|\gtrsim \Delta k\sim L_{1,\alpha}^{-1}$. Therefore the ansatz 
(\ref{eq19}) is valid for momenta with $|\sin k_\alpha|\gtrsim L_1^{-1/3}$ 
corresponding to energies not being close to the band center. 
Inserting (\ref{eq19}) in (\ref{eq18}) and choosing 
$x_\alpha\approx x_\beta$, we obtain
\begin{equation}
\label{eq20}
\Gamma_{\alpha\beta}\approx -2U^2\frac{1}{L_{1,\alpha}+L_{2,\alpha}}
\,\frac{\tilde g_0(k_\alpha+k_\beta)}{1-U\,\tilde g_0(k_\alpha+k_\beta)}
\end{equation}
giving rise to the Breit-Wigner width (for $E=0$)
\begin{equation}
\label{eq21}
\Gamma_{\alpha\beta}^{(1)}\approx
\frac{U^2/(4L_1)}{\sin^2 k_\alpha+\sin^2 k_\beta}\,
\frac{|\cos[(k_\alpha+k_\beta)/2]|}
{\cos^2[(k_\alpha+k_\beta)/2]+(U/4)^2}\ .
\end{equation}
These expressions differ by the $k$-dependent localization length from the 
result of Ref. \cite{jacquod2}. Since $\Gamma_{\alpha\beta}^{(1)}$ 
depends strongly on the momenta $k_\alpha$ and $k_\beta$, we determine 
the average with respect to these momenta \cite{jacquod2}~:
\begin{equation}
\label{eq22}
\Gamma=\frac{1}{4\pi^2}\int_{-\pi}^\pi dk_\alpha\,\int_{-\pi}^\pi dk_\beta\,
\delta(E-E_\alpha-E_\beta)\,/\nu_2(E)\ .
\end{equation}
Using Eq. (\ref{eq10}), relating $\Gamma$ with $L_2$, we can estimate 
for $E\to 0$ the two particle localization length as
\begin{eqnarray}
\nonumber
L_2&\approx& c(U)\,L_1^2\quad,\\
\label{eq23}
c(U)&=&
A\,\frac{2}{\pi^2}\int_0^\pi dk\frac{1}{(\sin^2 k+k_c^2)}
\,\frac{(U/4)^2}{\sin^2 k+(U/4)^2}\\
\nonumber 
&=& \frac{2\,A\,|U|}{\pi k_c}\,
\frac{s(\frac{U}{4})+k_c^2}
{|U|\,s(\frac{U}{4})\sqrt{s(k_c)}+4 k_c\,s(k_c)\,\sqrt{s(\frac{U}{4})}}
\end{eqnarray}
where $s(x)=1+x^2$, $A$ is a numerical prefactor and $k_c$ is a cutoff 
value to regularize the integral for small $k$ where the ballistic ansatz 
(\ref{eq19}) is invalid. For the values $k_c= 0.25$ and 
$A= 0.028$ the $U$-dependence of (\ref{eq23}) fits very well the 
numerical data for the slope $c(U)$ which can be seen in Fig. \ref{fig5}. 
We can considerably simplify the somewhat lengthy expression (\ref{eq23}) 
by neglecting the quadratic corrections $U^2$ and $k_c^2$ and 
slightly modifying $A$, 
\begin{equation}
\label{eq24}
c(U)\approx 0.074\frac{|U|}{|U|+1}\ .
\end{equation}
This approximation is numerically very accurate with an relative 
error smaller than $1\ \%$ for $0\le |U|\le 2.0$. 
The linear behavior of $c(U)$ for small $|U|$ is 
due to a combination of the logarithmic singularity in the density of states 
at $E\to 0$ and of large values of $\Gamma_{\alpha\beta}^{(1)}\sim U$ if 
$\cos[(k_\alpha+k_\beta)/2]\approx \pm U$. 

In view of the agreement between the numerical data and 
the theoretical estimate for $c(U)$, 
we conclude that the idea of diffusively moving particle pairs 
for short time scales finally becoming localized due to quantum interference
\cite{shep1,imry1,shep2,jacquod2} can indeed quantitatively explain the 
delocalization effect. For this it is important to evaluate carefully 
the Breit-Wigner width by taking into account the energy dependence of the 
one particle localization length. 

Despite this agreement 
we want to emphasize that the result (\ref{eq23}) is nevertheless based on 
several qualitative arguments. Actually, the application of the relation 
(\ref{eq10}) is somewhat problematic because both the Breit-Wigner width 
and the one-particle localization length do not have unique values due to 
their energy dependence. It is a priori not clear if the 
simple average (\ref{eq22}) is really sufficient and accurate. 
Furthermore, Eqs. (\ref{eq23}), (\ref{eq24}) depend on the artificial 
cutoff parameter $k_c$. Theoretically, we expect that $k_c\sim L_1^{-1/3}$ 
because of the invalidity of the ballistic ansatz (\ref{eq19}) for small 
momenta $|k|<k_c$. Numerically, the case $k_c=0.25$ indeed corresponds to 
$L_1\approx 50-100$ the largest considered $L_1$ values. However, the 
resulting dependence on $L_1$, i.~e. $L_2\sim L_1^{7/3}$ clearly disagrees 
with the numerical data. We attribute this to the fact that for 
small momenta according to $L_{1,\alpha}\approx L_1\sin^2 k_\alpha$ 
the effective size of the random hops is strongly reduced. This feature 
is not properly taken into account in the relation (\ref{eq10}). 
Therefore it would be interesting to carefully generalize this 
relation to the case where $\Gamma$ and $L_1$ 
have complicated distributions instead of unique values. 

\section{Conclusion} 

\label{sec4}

In this work, we have presented and applied two new accurate and efficient 
variants of the Greens function method originally introduced by von Oppen 
et al. \cite{oppen1} to study the TIP localization problem \cite{shep1}. Our 
results for the TIP localization length $L_2$ can be well fitted 
(Fig. \ref{fig5}) by the functional form $L_2=L_1[0.5+c(U)\,L_1]$. 
The behavior of the slope $c(U)\approx 0.074\,|U|/(1+|U|)$ 
is determined by the $U$-dependence of the Breit-Wigner width $\Gamma$ of 
non-interacting pair states. For this we presented an accurate estimate of 
$\Gamma$ extending former work of Jacquod et at. \cite{jacquod2}. 

We think that our results provide important additional evidence for the 
delocalization effect as such. In particular, we find for $U=2.0$ 
and $W=1.0$ an enhancement factor $L_2/(2\,L_1)\approx 11.5$. Our data 
is in qualitative agreement with former results of Song et al. 
\cite{song1}, who directly used the less efficient recursive Green function 
technique for smaller system sizes ($N\le 200$), and with very recent work 
\cite{leadbeater1,song2} based on the decimation method ($N\le 251$ and 
$N\le 300$). In view of this the original claim of R\"omer et al. 
\cite{roemer1} that there is no delocalization effect for infinite system 
size can no longer be maintained. To understand the transfer matrix data 
on which this claim was based, we remind that the considered disorder 
value $W=3.0$ was relatively large such that $L_2$ and $L_1$ are nearly 
equal. Using, 
the Green function method one can still measure the enhancement 
because $L_2>L_1/2$. However, in the transfer matrix approach there is 
a direct competition of $L_2$ with $L_1$ \cite{song2}. Furthermore, 
even for smaller disorder, the finite size behavior of the transmission 
eigenvalues is very subtle and one has to be careful about the 
finite size extrapolation here \cite{frahm5}. 

While the delocalization effect is now well established, the 
situation is less clear concerning the functional dependence of $L_2$ on 
$L_1$. The formerly observed powerlaw $L_2\sim L_1^\alpha$ with 
$\alpha\approx 1.45-1.65$ \cite{frahm1,song1,leadbeater1} was obtained 
by a fit ansatz without constant term. According to our above discussion 
(see Fig. \ref{fig4}) it is numerically not obvious to distinguish 
this powerlaw from the functional form we proposed above. Waintal et al. 
\cite{waintal1} gave an argument in favor of the former 
with $\alpha=1.5$. This argument is based on the multifractal properties 
of the interaction induced coupling matrix elements in combination 
with an estimate of $\Gamma$ using Fermi's golden rule. We believe 
that this analysis is indeed important and very relevant to the problem. 
Actually, our result (\ref{eq21}) for the Breit-Wigner width contains a 
strong dependence on the initial one-particle states due to partial 
momentum conservation. This leads to strong fluctuations of 
$\Gamma$ which are presumably directly related to the multifractal 
statistics of the coupling matrix elements. However, the analytical 
calculations of Ref. \cite{jacquod2} and of section \ref{sec3} clearly show 
that the simple application of Fermi's golden rule is not sufficient and 
the numerical data do not show the corresponding behavior $L_2\sim U^2$. 
To understand these issues in more detail further 
work is necessary. 

We emphasize that our numerical data and the estimate (\ref{eq23}) 
are valid for moderate interaction strengths $|U|\le 2.0$. 
For larger values of $U$ one expects that $L_2$ will be reduced 
due to the particular projector structure of the interaction operator 
$\hat U$. Waintal et al. \cite{waintal1} presented a duality transformation 
mapping the case of $|U|\gg 1$ to a similar problem with $|U|\ll 1$ 
and a different reference basis. According to this $L_2(U)$ should obey 
the duality relation $L_2(U)\approx L_2(\sqrt{24}/U)$ \cite{waintal1}. 

Finally, we mention that the numerical trick to evaluate efficiently 
the matrix $g_0$ via Eq. (\ref{eq6}) also works in higher 
dimensions, even though the gain is less spectacular. In $d$ dimensions 
and a system of total size (volume) $N$ one can calculate 
$G^{(1)}(E)$ by the recursive Green function method which provides 
an algorithm to evaluate (\ref{eq6}) with $N^{3+(d-1)/d}$ operations. 
In particular the case of two dimensions is important due to recent 
claims of Ortuno et al. and Cuevas \cite{ortuno1} for a delocalization 
transition for two interacting particles in $d=2$. We think it is necessary 
to consider larger system sizes as in Ref. \cite{ortuno1} in order 
to decide whether there is a real transition or a very strong delocalization 
with a finite but very large two-particle localization length as it 
was argued by Shepelyansky \cite{shep2}. 

\begin{acknowledgement}
The author thanks Dima Shepelyansky and Bertrand Georgeot for 
inspiring discussions. The Aspen Center for Physics is acknowledged 
for its hospitality during the workshop Quantum Chaos and Mesoscopic 
Systems at which a part of this work was done. 
\end{acknowledgement}


\end{document}